\documentclass[sigconf]{acmart}
\renewcommand\footnotetextcopyrightpermission[1]{} % removes footnote with conference information in first column
\setcopyright{none}
\usepackage{}
\usepackage{tikz}
\usepackage{amsmath}
\usepackage{comment}
\usepackage{booktabs} % For formal tables
\usepackage{xspace}
\usepackage{graphicx}
\usepackage{subfigure}
\usepackage{makecell}
\usepackage{diagbox}
\usepackage{algorithm}
\usepackage{algorithmic}
\usepackage{listings}
\usepackage{graphics}
\usepackage{epsfig}
\usepackage{color}
\usepackage{ifpdf}
\usepackage{mathrsfs}
\usepackage{algorithm}
\usepackage{algorithmic}
\usepackage{multirow}
\usepackage{verbatim}
\usepackage{pifont}
\usepackage{xcolor}
\usepackage{url}
\usepackage{color,xcolor}
\usepackage{array}
\usepackage{multirow}
\usepackage{bm}
\usepackage{subfigure}
\usepackage{algorithm}
\usepackage{xspace}
\usepackage{stmaryrd}
\usepackage{comment}
\usepackage{graphicx}
\usepackage{appendix}
\usepackage{hyperref}

\usepackage{amsfonts}
\usepackage{amsmath}

\usepackage{booktabs}
\usepackage{caption}
\usepackage{tikz}
\usepackage{xcolor}
\usepackage{listings,lstautogobble}
\usepackage{multirow}
\usepackage{syntax} 
\usepackage{xparse}
\usepackage{algorithm}
\usepackage{threeparttable}
\usepackage{proof}
\usepackage{enumitem}
\usepackage{tablefootnote}
\usepackage[normalem]{ulem}
\usepackage{fancyhdr}
 
\begin{document}
\thispagestyle{fancy}
% Set the page style to "fancy"...
%... then configure it.
\fancyhead{} % clear all header fields
\fancyhead[RO,LE]{\textbf{2023 Annual Computer Security Applications Conference (ACSAC)}}

\date{}

\title{\Large \bf Using Program Knowledge Graph to Uncover Software Vulnerabilities}

\author{\vspace{-0.2cm}Mengjie Xie}
\email{mengjix1@uci.edu}
\affiliation{
  \institution{University of California, Irvine}
  \country{}}

\author{\vspace{-0.2cm}Tamjid Al Rahat}
\email{tamjid@ucla.edu}
\affiliation{
    \institution{University of California, Los Angeles}
    \country{}}

\author{\vspace{-0.2cm}Wei Wang}
\email{weiwang@cs.ucla.edu}
\affiliation{
    \institution{University of California, Los Angeles}
    \country{}}
    
\author{\vspace{-0.2cm}Yuan Tian}
\email{yuant@ucla.edu}
\affiliation{
    \institution{University of California, Los Angeles}
    \country{}}

\begin{abstract}
In an increasingly interconnected and data-driven world, the importance of robust security measures cannot be overstated. A knowledge graph constructed with information extracted from the system along with the desired security behavior can be utilized to identify complex security vulnerabilities hidden underneath the systems. Unfortunately, existing security knowledge graphs are constructed from coarse-grained information extracted from publicly available vulnerability reports, which are not equipped to check actual security violations in real-world system implementations. In this poster, we present a novel approach of using \textit{Program Knowledge Graph} that is embedded with fine-grained execution information of the systems (e.g., callgraph, data-flow, etc.) along with information extracted from the public vulnerability and weakness datasets (e.g., CVE and CWE). We further demonstrate that our custom security knowledge graph can be checked against the standard queries generated by LLM, providing a powerful way to identify security vulnerabilities and weaknesses in critical systems.
\end{abstract}


\maketitle
\pagestyle{plain}
\thispagestyle{fancy}

%  array fingerprint
%  normal 改成 authentic

\newcounter{note}[section]
\newcommand{\frameName}{\textsc{ArrayID}\xspace}
\newcommand{\void}{\textsc{Void}\xspace}

   \renewcommand{\thenote}{\thesection.\arabic{note}}

\newif\ifrev
% comment out the following line after the revision is done
\revtrue
\ifrev

\def\UrlBreaks{\do\A\do\B\do\C\do\D\do\E\do\F\do\G\do\H\do\I\do\J
\do\K\do\L\do\M\do\N\do\O\do\P\do\Q\do\R\do\S\do\T\do\U\do\V
\do\W\do\X\do\Y\do\Z\do\[\do\\\do\]\do\^\do\_\do\`\do\a\do\b
\do\c\do\d\do\e\do\f\do\g\do\h\do\i\do\j\do\k\do\l\do\m\do\n
\do\o\do\p\do\q\do\r\do\s\do\t\do\u\do\v\do\w\do\x\do\y\do\z
\do\.\do\@\do\\\do\/\do\!\do\_\do\|\do\;\do\>\do\]\do\)\do\,
\do\?\do\'\do+\do\=\do\#}

\newcommand\tamjid[1]{{\color{blue}{\textbf{Tamjid:}{\em#1}}}}

\newcommand{\eg}{\emph{e.g.}\xspace}
\newcommand{\etal}{\emph{et al.}\xspace}
\newcommand{\etc}{\emph{etc}\xspace}
\newcommand{\ie}{\emph{i.e.}\xspace}

%\newcommand\yuan[1]{{\color{blue}{\textbf{Yuan:}{\em#1}}}}
%\newcommand\liam[1]{{\color{blue}{\textbf{Liam:}{\em#1}}}}
%\newcommand\tamjid[1]{{\color{blue}{\textbf{Tamjid:}{\em#1}}}}
%\newcommand\faysal[1]{{\color{blue}{\textbf{Faysal:}{\em#1}}}}
%\newcommand\wentao[1]{{\color{blue}{\textbf{Wentao:}{\em#1}}}}

%Turn Off comments 

%\newcommand\tamjid[1]{{\color{blue}{{#1}}}}
\newcommand\Mengjie[1]{{\color{green}{Mengjie:}{{#1}}}}

\newcommand{\toolname}{{TranSec}\xspace}

\lstset{
    language=C,
    basicstyle=\ttfamily\small,
    keywordstyle=\color{blue},
    commentstyle=\color{green!60!black},
    stringstyle=\color{red},
    showstringspaces=false,
    tabsize=4,
    breaklines=true,
    frame=single
}

% Define the Cypher language
\lstdefinelanguage{Cypher}{
    keywords={MATCH, WHERE, RETURN, CREATE, SET, DELETE, MERGE, ON, OPTIONAL, UNWIND, WITH, FOREACH},
    sensitive=true,
    morecomment=[l]{//},
    morecomment=[s]{/*}{*/},
    morestring=[b]",
    stringstyle=\color{blue},
    keywordstyle=\color{purple},
    morekeywords={MATCH, WHERE, RETURN, CREATE, SET, DELETE, MERGE, ON, OPTIONAL, UNWIND, WITH, FOREACH}
}

%------------------------------------------------------------------
\section{Introduction}
In the rapidly evolving landscape of security vulnerabilities, security analysts and researchers require novel approaches to establish connections between semantics of software vulnerabilities and their presence in real-world systems. The concept of a knowledge graph, initially introduced by Google, has found many applications in representing diverse forms of vulnerability information. However, its untapped potential for comprehensively capturing software defects and vulnerabilities within real systems has yet to be fully investigated.

Prior works in this domain are mostly focused on the development and management of knowledge graphs constructed from vulnerability databases~\cite{Qin}, open-source threat intelligence~\cite{gao2021system}, and cybersecurity threats~\cite{CSKB}.
Notably, while focusing on the extraction and integration of knowledge from diverse vulnerability sources, these works ignore the utilization of knowledge graphs to uncover real-world vulnerabilities in security-critical systems.

In this poster, we propose a novel approach of \textit{Program Knowledge Graph} that expands the vulnerability knowledge with program graphs (e.g., callgraph, data-flow, etc.) extracted from the implementation of security-critical systems. Further, leveraging prompt tuning with large language models (LLMs), we propose to automatically generate queries that assess the presence of specific vulnerabilities hidden underneath the complex systems. This methodology empowers developers to scrutinize and fortify their implementation against potential security risks at runtime. While existing work demands manual query construction, our approach seeks to alleviate this challenge by harnessing LLMs for query autogeneration.
\section{System Overview}
\begin{figure}
\centering
\includegraphics[width=0.39\textwidth]{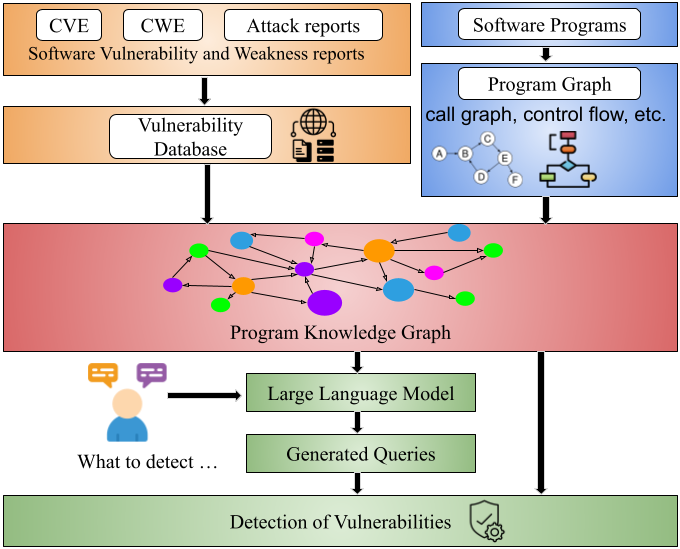}
\caption{Schematic architecture of our \textit{Program Knowledge Graph} to detect vulnerabilities in software programs.}
\label{fig:artchitecture-overview}
\vspace{-6mm}
\end{figure} 
The comprehensive architectural overview of the vulnerability detection process is presented in Figure \ref{fig:artchitecture-overview}. Our proposed approach encompasses a series of structured stages. First, known vulnerability data pertaining to Common Vulnerabilities and Exposures (CVE) and Common Weakness Enumeration (CWE) are methodically sourced from a vulnerability database and stored in a graph database. CWEs and CVEs offer insights into prevalent coding errors and identify vulnerable functions or methods, which allows security analysts to query vulnerability cases analogous to their code errors and assess associated security risks and potential impacts.

Secondly, software programs can be comprehensively analyzed and understood using graphs, where nodes represent program components, and edges depict relationships or flows between them. Program graphs, in this context, are a collection of such graphs that collectively represent the inner workings of a software program. In this poster, we have chosen call graphs as a prime example. However, our approach can easily be extended to other program graphs, such as control flow graphs, data flow graphs, and so on, to detect more complex code vulnerabilities. 

\begin{figure}
\centering
\includegraphics[width=0.37\textwidth]{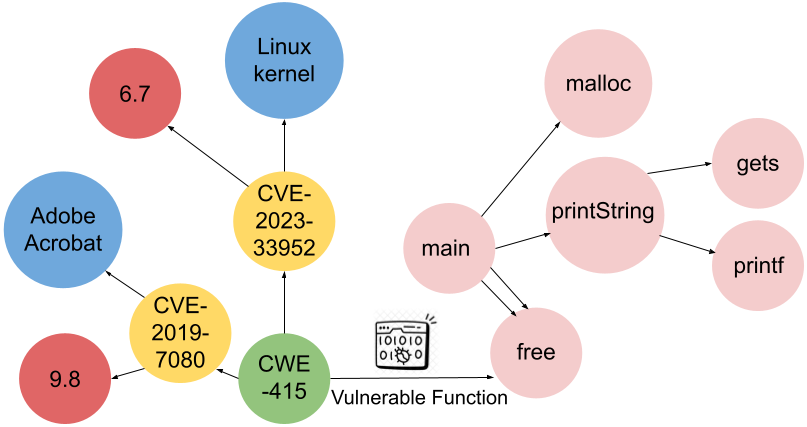}
\vspace{-2mm}
\caption{An example of \textit{Program Knowledge Graph} contains the call graph of the example C code in Figure \ref{fig:code-example} on the right side (colored in pink) and the security data displayed on the left side with CWE-415 pointing to a function call \textit{free} causing \textit{double free} vulnerability.}
\label{fig:KG-examples}
\vspace{-3mm}
\end{figure}

Next, \textit{Program Knowledge Graph} amalgamates vulnerability data and program graph, storing them in a graph database. Each node within this graph represents either a security data entity or a procedure call, with each edge signifying the relationships between source and destination nodes. Nodes and edges can also have attributes to store additional information, such as severity scores and affected products. 
An example of \textit{Program Knowledge Graph} is shown in Figure \ref{fig:KG-examples}, where nodes colored in pink show the call graph of the code example in Figure \ref{fig:code-example} while other nodes comprise the extracted vulnerability data. The arrows within the call graph symbolize calling relationships. \textit{Program  Knowledge Graph} creates additional edges connecting nodes from the vulnerability dataset to the vulnerable function calls extracted from the program.

% The green node CWE-415\cite{CWE-415} has the attribute "Program events" with value \textit{free} indicating that \textit{free} is a potential vulnerable function. When merging the security data and the call graph, \textit{Program  Knowledge Graph} creates an edge connecting node CWE-415 to the vulnerable function call \textit{free}. The yellow nodes are CVE instances categorized into CWE-415. Each CVE instance has CVSS3 scores colored in red and points to products it affects. 

Finally, we utilize LLM prompts to generate queries for checking vulnerabilities in the program. Precisely, developers and security analysts can use LLMs to generate complex queries pertinent to the intersection of vulnerability data and programs. This approach allows individuals without prior knowledge of any query language to provide LLMs with a prompt specifying the vulnerability to be checked, with LLMs subsequently generating the corresponding queries automatically. For example, Figure \ref{fig:cypher-query} showcases a Cypher query~\cite{cypher} that identifies the weakness of CWE-415 \textit{double free} in the C code shown in Figure \ref{fig:code-example}. Specifically, the query checks if a program calls \textit{free} twice on the same memory address, potentially corrupting the program’s memory management data structures.

\begin{figure}
    \centering
    \footnotesize
    \begin{lstlisting} [captionpos=b, frame=none]
void main() {
    char* ptr = (char*)malloc(8);
    printString(ptr);
    free(ptr);
    free(ptr);
}
void printString(char* ptr) {
    gets(ptr);
    printf(ptr);
}
    \end{lstlisting}
    \vspace{-5mm}
\caption{C code example with the vulnerability of calling \textit{free} twice on the same memory address.}
\label{fig:code-example}
\vspace{-4mm}
\end{figure}

\begin{figure}[h]
\centering
\footnotesize
\begin{lstlisting}[
    language=Cypher,
    captionpos=b,
    frame=none
]
MATCH (cwe:CWE {`CWE-ID`: "CWE-415"})
MATCH (node:CallGraph {Name: cwe.`events`})
WITH node.A AS arg1, COLLECT(node) AS sarg1
WITH arg1, sarg1, SIZE(sarg) AS nCount
WHERE nCount > 1
UNWIND sarg AS vulNodes
MATCH path=(CallGraph {Name: "main"})-[*]->(vulNodes)
RETURN path
\end{lstlisting} 
\vspace{-5mm}
\caption{Example of a Cypher query to detect a \textit{double free} vulnerability in the code example in Figure ~\ref{fig:code-example}.}
\label{fig:cypher-query}
\end{figure}

\section{Evaluation}
\begin{table}[H]
    \centering
    \footnotesize
    \begin{tabular}{|p{0.16\columnwidth}|p{0.32\columnwidth}|p{0.14\columnwidth}|p{0.11\columnwidth}|}
    \hline
        \textbf{CWE} & \textbf{Name} & \textbf{Program events} & \textbf{Code samples} \\
        \hline
        %CWE-134 & Use of Externally-Controlled Format String & printf, snprintf \\
        % \hline
        CWE-242 & Use of inherently dangerous function & gets, atoi, atol, atof & 4\\
        \hline
        %CWE-243 & Creation of chroot Jail Without Changing Working Directory & chdir \\
        % \hline
        CWE-401 & Missing Release of Memory after Effective Lifetime & malloc & 1\\
        \hline
        %CWE-401 & Missing Release of Memory after Effective Lifetime & free \\
        % \hline
        CWE-415 & Double free & free & 2\\
        \hline
        CWE-467 & Use of sizeof() on a pointer type & sizeof & 1\\
        \hline
        CWE-477 & Use of obsolete function & getpw, auto_ptr & 3\\
        \hline
        CWE-479 & Signal handler use of a non-reentrant function & syslog & 1\\
        \hline
        CWE-558 & Use of getlogin() in multithreaded application & getlogin & 1\\
        \hline
        %CWE-676 & Use of Potentially Dangerous Function & strcpy \\
        % \hline
        CWE-1341 & Multiple releases of same resource or handle & fclose, free & 2\\
    \hline
    \end{tabular}
    \caption{Our queries for the \textit{Program Knowledge Graph} successfully identified 14 out of 15 weaknesses in the code samples we collected from 8 categories of CWEs. }
    \label{Tab:detectable-CWEs}
    \vspace{-6mm}
\end{table}

To evaluate our proposed method, we collected 15 code examples in C/C++ as benchmarks from CWE sites associated with 8 common software weaknesses, as shown in Table \ref{Tab:detectable-CWEs}. For our evaluation, we leverage Neo4j, a popular open-source graph database, as our backend for graph traversal. We provide the CSV formatted dataset of CVEs and CWEs, along with the call graph constructed for the collected code examples. By utilizing Cypher, a widely adopted query language in both industrial and research settings~\cite{cypher}, we successfully identified the weakness of 14 out of 15 benchmarks. We could not identify the weakness of CWE-401, as it requires the inclusion of data flow graphs within the program graph for effective detection, as call graphs alone may not suffice. Prior works in security knowledge graphs primarily emphasized the design and integration of vulnerability data rather than detection, making an apple-to-apple comparison infeasible. We plan to expand our approach to include various graph types, such as data flow graphs, to detect more complex vulnerabilities and extend our evaluation by comparing them with existing vulnerability detection methods.

% In this poster,  The Neo4j database takes in CSV format files to construct nodes, edges, and attributes, and Neo4j’s built-in Cypher query language simplifies . Twelve CWE vulnerabilities have been tested, and meticulously crafted queries successfully identified total 20 defenseless code snippets sourced from corresponding official CWE sites.  The "Program events" column in the table enumerates vulnerable functions or methods. These CWEs are sourced from the CWE VIEW: Weaknesses in Software Written in C++\cite{CWE-659}. However, CWEs like CWE-401: \textit{Missing Release of Memory after Effective Lifetime}\cite{CWE-401}, may require the inclusion of data flow graphs within the program graph for effective detection, as call graphs alone may not suffice.

%\input{02-background}
%\input{04-analysis}
%\input{05-experiments}
\section{Conclusion}
We propose the concept of \textit{Program Knowledge Graph} by integrating program graph and security data and subsequently auto-generating queries by leveraging LLM's prompt tuning to discover vulnerabilities within software code. We are going to continue working on this research project to further investigate the systematic methodology of automating the query generation process by leveraging LLMs.

% \small
{\footnotesize
\bibliographystyle{acm}
\bibliography{reference}
}

\end{document}